\documentclass[%
 aip,
 amsmath,amssymb,
 preprint,%
]{revtex4-1}
\usepackage{amscd}
\usepackage{amsmath}
\usepackage{mathrsfs}
\usepackage{color}
\usepackage{graphicx}
\usepackage{dcolumn}
\usepackage{bm}

\newtheorem{prop}{Proposition}

\allowdisplaybreaks

\begin{document}

\title[]{The transformations between $N=2$ supersymmetric KdV and HD equations}

\date{\today}
\author{Kai Tian \text{and}  Q.~P.~Liu}
\affiliation{Department of Mathematics, China University of Mining
and Technology, Beijing 100083, P. R. China}

\begin{abstract}
The $N=2$ supercomformal transformations are employed to study
supersymmetric integrable systems. It is proved that two known $N=2$
supersymmetric Harry Dym equations are transformed into two $N=2$
supersymmetric modified Kortweg-de Vries equations, thus are
connected with two $N=2$ supersymmetric Kortweg-de Vries equations.
\end{abstract}


\maketitle


\section{Introduction}
The reciprocal transformation, also known as hodograph transformation, plays an important role when we investigate relations among some nonlinear evolution equations. For instance, the Harry Dym (HD) equation (or hierarchy) \cite{calogero}, which is 
invariant under a kind of reciprocal transformation, is also reciprocally linked to the Korteweg-de Vries (KdV) equation (or hierarchy). The Camassa-Holm equation is shown to be linked to the first negative flow of the KdV hierarchy \cite{fuch}. The Kawamoto equation is transformed to the common modification of the Sawada-Kotera and Kaup-Kupershmidt equations by reciprocal transformation \cite{kawa} and many other examples exist.
Apart from its own interest, the reciprocal transformation is also a powerful tool to investigate integrable properties of nonlinear evolution equations. Indeed, recursion operators, bi-Hamiltonian structures and  solutions of one given  equation could be constructed from the corresponding properties of the other equation through the associated  reciprocal transformation. With the help of the reciprocal transformation, the abundant symmetry structures of Kawamoto-type equations and Harry Dym equation were revealed \cite{carillo}.

Very recently, the reciprocal transformation was generalized to $N=1$ supersymmetric equations \cite{liu}, where a general procedure to construct supersymmetric reciprocal transformation was presented. As applications, one of the supersymmetric HD equations was shown to be reciprocally linked to the supersymmetric modified KdV equation, and the supersymmetric Kawamoto equation, as a fifth order analog of HD equation, was transformed to the supersymmetric modified Sawada-Kotera equation. As in the classical case, the supersymmetric reciprocal transformation could be employed to explore integrable properties of supersymmetric equations, which was illustrated by constructing the recursion operators and bi-Hamiltonian structures of the supersymmetric HD and Kawamoto equations. More examples of supersymmetric reciprocal transformation can be found in Ref. \onlinecite{liu2}.

Besides $N=1$ supersymmetric generalizations, the integrable systems also admit $N=2$ extended supersymmetric generalizations. The idea could almost be traced back to the usage of the supersymmetry in the quantum field theory \cite{kulish}. As a striking feature, $N=2$ extended case distinguishes itself from $N=1$ non-extended case by the possibility to supply new classical integrable systems.
The $N=2$ supersymmetric KdV equations were proposed more than twenty years ago \cite{laberge,labelle} and have been studied extensively since then. Various results  for these equations  have been obtained, including Lax representations \cite{labelle,popo}, bi-Hamiltonian structures \cite{oevel}, bilinear formulism \cite{zhang} and so on. In addition to the KdV equation, some other integrable equations, such as HD equation, Camassa-Holm equation were also generalized to the $N=2$ super space \cite{brunelli,popo2} (see also Refs. \onlinecite{araytn} and \onlinecite{ll}).

With the success of establishing the $N=1$ supersymmetric reciprocal transformation, it would be important to extend the results to the $N=2$ spersymmetric case and to study the possible links between $N=2$ supersymmetric HD and KdV equations. In this paper, we will show that such extension is indeed possible. We will show that the supercomformal transformations serve our need. The paper is organized as follows. In the next section, we recall the $N=2$ supersymmetric KdV and HD equations. In the section three, the $N=2$ supercomformal transformations and the relevant results will be reviewed and elaborated. Then in the section four, the transformations between the two $N=2$ supersymmetric HD equations and supersymmetric KdV equations are given.
Final section contains some comments and open questions.

\section{$N=2$ Supersymmetric KdV and HD equations}
The $N=2$ supersymmetric KdV equation is given by the one-parameter system
\begin{equation}\label{kdva}
\Phi_{\tau}=\frac{1}{4}\Big(\Phi_{3y}-3[\Phi(\mathbb{D}_1\mathbb{D}_2\Phi)]_y-\frac{a-1}{2}[\mathbb{D}_1\mathbb{D}_2\Phi^2]_y-3a\Phi^2\Phi_y\Big),
\end{equation}
which is denoted by  SKdV$_a$ equation in literature. Some conventional criteria have been adopted to study its integrability and the system is known to be integrable only for certain values of the parameter $a$. In fact, the existence of infinitely many conservation laws implies that $a$ only takes three values, $-2$, $1$ and $4$ .\cite{labelle} The singularity analysis   also leads to  the same conclusion \cite{bourque}. For these three cases, Eq. \eqref{kdva} was shown to admit Lax representations \cite{labelle,popo}
\begin{align*}
a=4: && \mathbb{L}_4=-(\mathbb{D}_1\mathbb{D}_2+\Phi)^2, && \frac{\partial}{\partial \tau_n}\mathbb{L}_4=[(\hat{\mathbb{L}}_4^n\mathbb{L}_4^{1\over2})_{\geq 0},\mathbb{L}_4],\\
a=-2: && \mathbb{L}_{-2}=\partial_y^2+\mathbb{D}_1\Phi\mathbb{D}_2-\mathbb{D}_2\Phi\mathbb{D}_1, && \frac{\partial}{\partial \tau_n}\mathbb{L}_{-2}=[(\mathbb{L}_{-2}^{n\over2})_{\geq 0},\mathbb{L}_{-2}],\\
a=1: && \mathbb{L}_1=-\partial_y^{-1}\mathbb{D}_1\mathbb{D}_2(\mathbb{D}_1\mathbb{D}_2+\Phi), && \frac{\partial}{\partial \tau_n}\mathbb{L}_{1}=[(\mathbb{L}_{1}^{3n})_{\geq 1},\mathbb{L}_{1}],
\end{align*}
where we are using the standard notations and the subscripts $_{\geq  k}$ denote the corresponding projections. It is remarked that that the operator $\mathbb{L}_4$ possesses two different square roots, namely
\begin{equation*}
\hat{\mathbb{L}}_4=i(\mathbb{D}_1\mathbb{D}_2+\Phi),\qquad\mbox{and}\qquad \mathbb{L}_4^{1\over2}=\partial_y+\cdots.
\end{equation*}
The non-uniqueness of roots of Lax operator results in the SKdV${_4}$ equation which admits twice as many conserved quantities as those of the other two systems \cite{oevel}. It is easy to see, from the Lax representation, that the SKdV${_4}$ equation is not the first non-trivial flow in its hierarchy, while the first one is
\begin{equation}\label{kdv41}
\Phi_{\tau_{1}}=\frac{1}{2}(\mathbb{D}_1\mathbb{D}_2\Phi_y)+2\Phi\Phi_y.
\end{equation}

Let us now turn to the Harry Dym case. By considering the most general $N=2$ super Lax operator which consists of  differential operators only, two different $N=2$ supersymmetric Harry Dym hierarchies were presented \cite{brunelli}. One is formulated as
\begin{equation}\label{eq21}
\frac{\partial}{\partial t_n}L_1=\Big[\big(\hat{L}_1^n L_1^{1\over 2}\big)_{\geq 2},L_1\Big], \quad n=0,1,2.\cdots
\end{equation}
where the Lax operator is
\begin{equation}\label{eq22}
L_1=-(W\mathcal{D}_1\mathcal{D}_2)^2,
\end{equation}
whose two different square roots are given by
\begin{align*}
\hat{L}_1= & iW\mathcal{D}_1\mathcal{D}_2,\\
L_1^{1\over 2}= & W\partial_x+\frac{1}{2}\Big[(\mathcal{D}_1W)\mathcal{D}_1+(\mathcal{D}_2W)\mathcal{D}_2-W_x\Big]\\
& +\frac{1}{4}\Big[-2(\mathcal{D}_1\mathcal{D}_2W)-(\mathcal{D}_2W)(\mathcal{D}_1W)W^{-1}\Big]\mathcal{D}_1\mathcal{D}_2\partial_x^{-1}+\cdots\cdots,
\end{align*}
respectively. Based on the Lax equation \eqref{eq21}, it is easy to write down the first two non-trivial flows in this hierarchy explicitly, namely,
\begin{align}
W_{t_1}= & \frac{i}{2}(\mathcal{D}_1\mathcal{D}_2W_x)W^2,\label{hdt1}\\
W_{t_2}= & \frac{1}{8}\Big(2W_{3x}W^3-6(\mathcal{D}_1\mathcal{D}_2W_x)(\mathcal{D}_1\mathcal{D}_2W)W^2
-3(\mathcal{D}_2W_{2x})(\mathcal{D}_2W)W^2 \nonumber\\
&~~~~~-3(\mathcal{D}_1W_{2x})(\mathcal{D}_1W)W^2\Big).\label{hd}
\end{align}

The other $N=2$ supersymmetric Harry Dym hierarchy is defined as
\begin{equation}\label{eq23}
\frac{\partial}{\partial t_n}L_2=\Big[\big(L_2^{n\over 2}\big)_{\geq 2},L_2\Big]
\end{equation}
with the Lax operator
\begin{equation}\label{eq24}
L_2=\frac{1}{2}(\mathcal{D}_1W^2\mathcal{D}_1+\mathcal{D}_2W^2\mathcal{D}_2)\partial_x.
\end{equation}
The first non-trivial flow of this hierarchy reads as
\begin{align}
W_{t_3}=&\frac{1}{8}\Big(2W_{3x}W^3-3(\mathcal{D}_2W_{2x})(\mathcal{D}_2W)W^2-3(\mathcal{D}_1W_{2x})(\mathcal{D}_1W)W^2 \nonumber \\
&~~~+3(\mathcal{D}_2W)(\mathcal{D}_1W)(\mathcal{D}_1\mathcal{D}_2W_x)W\Big).\label{eq25}
\end{align}

In both cases, $W=W(x,\theta_1,\theta_2,t)$ is a bosonic super field.

\section{The Superconformal Transformation of $N=2$ Super Space}

A super diffeomorphism between two super spaces is named as a
superconformal transformation providing the super derivatives are
transformed covariantly (see Ref. \onlinecite{mathieu} and the references
there). Let us consider the super diffeomorphism between  the super
spaces  $(y,\varrho_1,\varrho_2)$ and  $(x,\theta_1,\theta_2)$
\begin{equation}\label{eq11}
\begin{aligned}
y \rightarrow & x=x(y,\varrho_1,\varrho_2),\\
\varrho_1 \rightarrow & \theta_1=\theta_1(y,\varrho_1,\varrho_2),\\
\varrho_2 \rightarrow & \theta_2=\theta_2(y,\varrho_1,\varrho_2).
\end{aligned}
\end{equation}
The associated super derivatives for each super space are
respectively denoted by
\[
\mathbb{D}_k=\partial_{\varrho_k}+{\varrho_k}\partial_y,
\qquad\mathcal{D}_k=\partial_{\theta_k}+{\theta_k}\partial_x,\quad
(k=1, 2).
\]

For the super diffeomorphism \eqref{eq11}, we have the following relations
\addtocounter{equation}{1}
\begin{align}
\mathbb{D}_1 = \Big((\mathbb{D}_1x)-\theta_1(\mathbb{D}_1\theta_1)-\theta_2(\mathbb{D}_1\theta_2)\Big)\frac{\partial}{\partial x} + (\mathbb{D}_1\theta_1)\mathcal{D}_1 + (\mathbb{D}_1\theta_2)\mathcal{D}_2 \tag{\theequation a},\\
\mathbb{D}_2 =
\Big((\mathbb{D}_2x)-\theta_1(\mathbb{D}_2\theta_1)-\theta_2(\mathbb{D}_2\theta_2)\Big)\frac{\partial}{\partial
x} + (\mathbb{D}_2\theta_1)\mathcal{D}_1 +
(\mathbb{D}_2\theta_2)\mathcal{D}_2 \tag{\theequation b},
\end{align}
hence, to ensure the super derivatives transforming covariantly,
i.e.
\begin{equation}\label{eq15}
\mathbb{D}_1  = (\mathbb{D}_1\theta_1)\mathcal{D}_1 +
(\mathbb{D}_1\theta_2)\mathcal{D}_2,\quad
\mathbb{D}_2  =
(\mathbb{D}_2\theta_1)\mathcal{D}_1 +
(\mathbb{D}_2\theta_2)\mathcal{D}_2,
\end{equation}
the super diffeomorphism \eqref{eq11} must be constrained by
\begin{equation}\label{eq14}
(\mathbb{D}_1x)  =
\theta_1(\mathbb{D}_1\theta_1)+\theta_2(\mathbb{D}_1\theta_2),\quad
(\mathbb{D}_2x)  =
\theta_1(\mathbb{D}_2\theta_1)+\theta_2(\mathbb{D}_2\theta_2).
\end{equation}

However,  the constraints \eqref{eq14} are not sufficient to ensure
the super diffeomorphism to be superconformal. Some further
constraints are resulted from the relation
\[
\mathcal{D}_1^2=\mathcal{D}_2^2=\partial_x.
\]
Through cumbersome, otherwise straightforward calculation, one
obtains
\begin{equation}\label{eq19}
(\mathbb{D}_1\theta_2)  = -(\mathbb{D}_2\theta_1), \quad
(\mathbb{D}_2\theta_2)  = (\mathbb{D}_1\theta_1).
\end{equation}

Summing up, we emphasize that under the constraints \eqref{eq14} and
\eqref{eq19}, the super diffeomorphism \eqref{eq11} is a
superconformal transformation, which can be formulated by
\addtocounter{equation}{1}
\begin{align}
\mathcal{D}_1 & = K^{-1}\Big((\mathbb{D}_1\theta_1)\mathbb{D}_1+(\mathbb{D}_2\theta_1)\mathbb{D}_2\Big)  \tag{\theequation a},\label{eq18a}\\
\mathcal{D}_2 & =
K^{-1}\Big(-(\mathbb{D}_2\theta_1)\mathbb{D}_1+(\mathbb{D}_1\theta_1)\mathbb{D}_2\Big)
\tag{\theequation b},\label{eq18b}
\end{align}
where $K=(\mathbb{D}_1\theta_1)^2+(\mathbb{D}_2\theta_1)^2$.

The following formulas, which can be checked directly,
\begin{align}
\mathcal{D}_1\mathcal{D}_2 &= K^{-1}\left[\mathbb{D}_1\mathbb{D}_2+\frac{1}{2}(\mathbb{D}_2\log K)\mathbb{D}_1-\frac{1}{2}(\mathbb{D}_1\log K)\mathbb{D}_2\right], \label{eq16}\\
\partial_x &= K^{-1}\left[\partial_y-\frac{1}{2}(\mathbb{D}_2\log K)\mathbb{D}_2-\frac{1}{2}(\mathbb{D}_1\log
K)\mathbb{D}_1\right],
\label{eq17}
\end{align}
are very useful when we study the link between the $N=2$
supersymmetric HD hierarchies to supersymmetric MKdV type
hierarchies.
\section{Reciprocal transformations}
In this section, we show that both $N=2$ supersymmetric HD
hierarchies mentioned in the section two are linked with
supersymmetric MKdV type hierarchies via superconformal
transformations. First, let us consider the behavior of the Lax
operators $L_1$ and $L_2$ under certain superconformal
transformations.

For the Lax operator $L_1=-(W\mathcal{D}_1\mathcal{D}_2)^2$, if we assume $W=K$, then according to the formula \eqref{eq16}, we have the new operator in the super space $(y,\varrho_1,\varrho_2)$
\begin{equation}\label{eq31}
L_{1m}=-\left[\mathbb{D}_1\mathbb{D}_2+\frac{1}{2}(\mathbb{D}_2\log K)\mathbb{D}_1-\frac{1}{2}(\mathbb{D}_1\log K)\mathbb{D}_2\right]^2.
\end{equation}
Let $U=1/2\log K$, then
\begin{equation}\label{eq32}
L_{1m}=-\left[\mathbb{D}_1\mathbb{D}_2+(\mathbb{D}_2U)\mathbb{D}_1-(\mathbb{D}_1U)\mathbb{D}_2\right]^2.
\end{equation}
Applying the gauge transformation on $L_{1m}$, we have
\[
e^{-U}L^{1m}e^U=-[\mathbb{D}_1\mathbb{D}_2+(\mathbb{D}_1\mathbb{D}_2U)+(\mathbb{D}_2U)(\mathbb{D}_1U)]^2 \equiv -[\mathbb{D}_1\mathbb{D}_2+\Phi]^2
\]
which is nothing but the Lax operator $\mathbb{L}_4$ of SKdV$_4$
equation. This implies that $L_{1m}$ should be a Lax operator for
the modification of SKdV$_4$ equation. In fact, we have
\begin{prop}
The Lax equation
\begin{equation}\label{eq34}
\frac{\partial}{\partial
\tau_n}L_{1m}=\Big[\big(\hat{L}_{1m}^nL_{1m}^{1\over2}\big)_{\geq
1}, L_{1m}\Big], \qquad n=0,1,2,\cdots
\end{equation}
defines a $N=2$ supersymmetric hierarchy, where
\begin{align*}
\hat{L}_{1m} = i\left[\mathbb{D}_1\mathbb{D}_2+(\mathbb{D}_2U)\mathbb{D}_1-(\mathbb{D}_1U)\mathbb{D}_2\right],\quad\mbox{and}\quad
L_{1m}^{1/2} = \partial_y + \cdots\cdots.
\end{align*}

The first two non-trivial flows in this hierarchy are explicitly given by
\begin{align}
U_{\tau_1} =& \frac{i}{2}\Big[(\mathbb{D}_1\mathbb{D}_2U_y) + 2(\mathbb{D}_1\mathbb{D}_2U)U_y + 2(\mathbb{D}_2U_y)(\mathbb{D}_1U) + 2(\mathbb{D}_2U)(\mathbb{D}_1U)\Big]\label{eq35},\\
U_{\tau_2} =& \frac{1}{4}\Big[U_{3y} - 2U_y^3 - 6(\mathbb{D}_1\mathbb{D}_2U_y)(\mathbb{D}_1\mathbb{D}_2U) - 6(\mathbb{D}_1\mathbb{D}_2U)^2U_y - 3(\mathbb{D}_2U_y)(\mathbb{D}_2U)U_y \nonumber\\
&~~~ - 9(\mathbb{D}_2U_y)(\mathbb{D}_1U)(\mathbb{D}_1\mathbb{D}_2U) - 9(\mathbb{D}_2U)(\mathbb{D}_1U_y)(\mathbb{D}_1\mathbb{D}_2U) \nonumber\\
&~~~ - 6(\mathbb{D}_2U)(\mathbb{D}_1U)(\mathbb{D}_1\mathbb{D}_2U_y)
- 3(\mathbb{D}_1U_y)(\mathbb{D}_1U)U_y\Big].\label{eq36}
\end{align}
Furthermore, the $\tau_2$ flow \eqref{eq36} is the modification of
SKdV$_4$ equation with the Miura transformation
\[
\Phi=(\mathbb{D}_1\mathbb{D}_2U)+(\mathbb{D}_2U)(\mathbb{D}_1U).
\]
\end{prop}

\noindent {\em Proof}: Direct calculations.

The superconformal transformation provides us a link between the
spatial variables. To have a complete picture, we need to find the
counterpart for the temporal variables.  For the couple of Lax
hierarchies \eqref{eq21} and \eqref{eq34}, the relations are given
by
\begin{equation}
\frac{\partial}{\partial t_n}-\big(\hat{L}_1^n L_1^{1\over
2}\big)_{\geq 2} =
\frac{\partial}{\partial\tau_n}-\big(\hat{L}_{1m}^nL_{1m}^{1\over2}\big)_{\geq
1}.
\end{equation}
We now calculate the explicit transformations for the first and
second flows. In the simplest case, namely  $n=1$, since
\begin{align*}
\big(\hat{L}_1 L_1^{1\over 2}\big)_{\geq 2} =& iW^2\partial_x\mathcal{D}_1\mathcal{D}_2 + \frac{i}{2}(\mathcal{D}_1W)W\partial_x\mathcal{D}_2 - \frac{i}{2}(\mathcal{D}_2W)W\partial_x\mathcal{D}_1 + \frac{i}{2}W\mathcal{D}_1\mathcal{D}_2 \\
=& i\partial_y\mathbb{D}_1\mathbb{D}_2 - i(\mathbb{D}_1U)\partial_y\mathbb{D}_2 + i(\mathbb{D}_2U)\partial_y\mathbb{D}_1 -iU_y\mathbb{D}_1\mathbb{D}_2 \\
&+ \Big(-i(\mathbb{D}_1U_y)+i(\mathbb{D}_1U)U_y\Big)\mathbb{D}_2 + \Big(i(\mathbb{D}_2U_y)-i(\mathbb{D}_2U)U_y\Big)\mathbb{D}_1,\\
\big(\hat{L}_{1m}L_{1m}^{1/2}\big)_{\geq 1} =& i\partial_y\mathbb{D}_1\mathbb{D}_2 - i(\mathbb{D}_1U)\partial_y\mathbb{D}_2 + i(\mathbb{D}_2U)\mathbb{D}_1 + i\Big((\mathbb{D}_1\mathbb{D}_2U)+(\mathbb{D}_2U)(\mathbb{D}_1U)\Big)\partial_y - i U_y\mathbb{D}_1\mathbb{D}_2 \\
&+ \frac{1}{2}\Big(-3i(\mathbb{D}_1U_y)+3i(\mathbb{D}_1U)U_y+i(\mathbb{D}_2U)(\mathbb{D}_1\mathbb{D}_2U)\Big)\mathbb{D}_2 \\
&+
\frac{1}{2}\Big(3i(\mathbb{D}_2U_y)-3i(\mathbb{D}_2U)U_y+i(\mathbb{D}_1U)(\mathbb{D}_1\mathbb{D}_2U)\Big)\mathbb{D}_1.
\end{align*}
Hence, we have
\begin{align*}
\frac{\partial}{\partial t_1} =& \frac{\partial}{\partial \tau_1} - i\Big((\mathbb{D}_1\mathbb{D}_2U)+(\mathbb{D}_2U)(\mathbb{D}_1U)\Big)\partial_y + \frac{i}{2}\Big((\mathbb{D}_1U_y)-(\mathbb{D}_1U)U_y-(\mathbb{D}_2U)(\mathbb{D}_1\mathbb{D}_2U)\Big)\mathbb{D}_2 \\
&+
\frac{i}{2}\Big(-(\mathbb{D}_2U_y)+(\mathbb{D}_2U)U_y-(\mathbb{D}_1U)(\mathbb{D}_1\mathbb{D}_2U)\Big)\mathbb{D}_1.
\end{align*}

Similarly, when $n=2$, we have
\begin{align*}
\frac{\partial}{\partial t_2} =& \frac{\partial}{\partial \tau_2}+\big(\hat{L}_1^2 L_1^{1\over 2}\big)_{\geq 2}-\big(\hat{L}_{1m}^2L_{1m}^{1\over2}\big)_{\geq 1}\\
=& \frac{\partial}{\partial \tau_2} + \frac{1}{2}\Big(3(\mathbb{D}_1\mathbb{D}_2U)^2-U_{2y}+U_y^2+(\mathbb{D}_2U_y)(\mathbb{D}_2U) + 4(\mathbb{D}_2U)(\mathbb{D}_1U)(\mathbb{D}_1\mathbb{D}_2U)\\
&\quad\quad\quad +(\mathbb{D}_1U_y)(\mathbb{D}_1U)\Big)\partial_y\\
&\quad +\frac{1}{4}\Big(-(\mathbb{D}_2U_{2y})+(\mathbb{D}_2U_y)U_y+4(\mathbb{D}_2U)(\mathbb{D}_1\mathbb{D}_2U)^2 + (\mathbb{D}_2U)U_{2y}+4(\mathbb{D}_2U)(\mathbb{D}_1U_y)(\mathbb{D}_1U)\\
&\quad\quad\quad - 5(\mathbb{D}_1U_y)(\mathbb{D}_1\mathbb{D}_2U) - (\mathbb{D}_1U)(\mathbb{D}_1\mathbb{D}_2U_y) + 4(\mathbb{D}_1U)(\mathbb{D}_1\mathbb{D}_2U)U_y\Big)\mathbb{D}_2\\
&\quad + \frac{1}{4}\Big(-(\mathbb{D}_1U_{2y})+(\mathbb{D}_1U_y)U_y+4(\mathbb{D}_1U)(\mathbb{D}_1\mathbb{D}_2U)^2+(\mathbb{D}_1U)U_{2y} + 4 (\mathbb{D}_2U_y)(\mathbb{D}_2U)(\mathbb{D}_1U) \\
&\quad\quad\quad +
5(\mathbb{D}_2U_y)(\mathbb{D}_1\mathbb{D}_2U)+(\mathbb{D}_2U)(\mathbb{D}_1\mathbb{D}_2U_y)
- 4(\mathbb{D}_2U)(\mathbb{D}_1\mathbb{D}_2U)U_y\Big)\mathbb{D}_1.
\end{align*}

Let us now consider the other $N=2$ supersymmetric Harry Dym
equation, whose  Lax operator is
$L_2=1/2(\mathcal{D}_1W^2\mathcal{D}_1+\mathcal{D}_2W^2\mathcal{D}_2)\partial_x$.
In this case,  we also assume $W=K$. From the formulas \eqref{eq18a}
\eqref{eq18b}, we have
\begin{equation*}
\frac{1}{2}(\mathcal{D}_1W^2\mathcal{D}_1+\mathcal{D}_2W^2\mathcal{D}_2)
=
K\partial_y+\frac{1}{2}(\mathbb{D}_2K)\mathbb{D}_2+\frac{1}{2}(\mathbb{D}_1K)\mathbb{D}_1.
\end{equation*}
Then taking the formula \eqref{eq17} into account, $L_2$ is transformed to
\begin{align}\label{eq37}
L_{2m} =& \left[ K\partial_y+\frac{1}{2}(\mathbb{D}_2K)\mathbb{D}_2+\frac{1}{2}(\mathbb{D}_1K)\mathbb{D}_1\right] K^{-1}\left[\partial_y-\frac{1}{2}(\mathbb{D}_2\log K)\mathbb{D}_2-\frac{1}{2}(\mathbb{D}_1\log K)\mathbb{D}_1\right] \nonumber\\
=&
\partial_y^2-U_y\partial_y-\frac{1}{2}(\mathbb{D}_2U)(\mathbb{D}_1U)\mathbb{D}_1\mathbb{D}_2
+\frac{1}{4}\Big(-2(\mathbb{D}_2U_y)+(\mathbb{D}_2U)U_y-(\mathbb{D}_1U)(\mathbb{D}_1\mathbb{D}_2U)\Big)\mathbb{D}_2 \nonumber\\
&
+\frac{1}{4}\Big(-2(\mathbb{D}_1U_y)+(\mathbb{D}_1U)U_y+(\mathbb{D}_2U)(\mathbb{D}_1\mathbb{D}_2U)\Big)\mathbb{D}_1.
\end{align}
By direct calculation, we have
\begin{prop}
The Lax equation
\begin{equation}\label{eq38}
\frac{\partial}{\partial
\tau_n}L_{2m}=\Big[\big(L_{2m}^{n\over2}\big)_{\geq
1},L_{2m}\Big],\qquad n=1,3,\cdots
\end{equation}
defines a $N=2$ supersymmetric hierarchy, whose first non-trivial flow is
\begin{align}\label{eq39}
U_{\tau_3} =& \frac{1}{16}\Big(4U_{3y} - 2U_y^3 - 3(\mathbb{D}_2U_y)(\mathbb{D}_2U)U_y + 3(\mathbb{D}_2U_y)(\mathbb{D}_1U)(\mathbb{D}_1\mathbb{D}_2U) \nonumber\\
 &~~~~~ + 3(\mathbb{D}_2U)(\mathbb{D}_1U_y)(\mathbb{D}_1\mathbb{D}_2U)-
 3(\mathbb{D}_1U_y)(\mathbb{D}_1U)U_y\Big).
\end{align}
Furthermore, the equation \eqref{eq39} is a modification of the
SKdV$_{-2}$ equation with the Miura type transformation
\begin{equation}\label{miu}
\Phi=\frac{1}{2}(\mathbb{D}_1\mathbb{D}_2U)+\frac{1}{4}(\mathbb{D}_2U)(\mathbb{D}_1U).
\end{equation}
\end{prop}

\noindent {\em Proof}: Direct calculations.

\noindent {\bf Remark:} 
As a byproduct of above
results, a relation could be inferred between the bosonic limit of
SKdV$_{-2}$ and that of Eq. \eqref{eq39}. Let
$\Phi=\phi_0+\theta_2\theta_1\phi_1$ and
$U=u_0+\theta_2\theta_1u_1$, then the bosonic limits of SKdV$_{-2}$
and Eq. \eqref{eq39} are respectively given by
\[
\begin{aligned}
\phi_{0,\tau_3} =& \frac{1}{4}\Big(\phi_{0,3y}+6\phi_{0,y}\phi_0^2\Big),\\
\phi_{1,\tau_3} =&
\frac{1}{4}\Big(\phi_{1,3y}-6\phi_{1,y}\phi_1+6\phi_{1,y}\phi_0^2+12\phi_1\phi_{0,y}\phi_0-6\phi_{0,2y}\phi_{0,y}\Big),
\end{aligned}
\]
and
\[
\begin{aligned}
u_{0,\tau_3} =& \frac{1}{8}\Big(2u_{0,3y}-u_{0,y}^3\Big),\\
u_{1,\tau_3} =& \frac{1}{8}\Big(2u_{1,3y}+3u_{1,y}u_1^2\Big).
\end{aligned}
\]
The Miura transformation between them is
\[
\phi_0 =  \frac{1}{2}u_1,\quad \phi_1 =
\frac{1}{4}\Big(-2u_{0,2y}+u_{0,y}^2+u_1^2\Big).
\]


As above, we derive the transformation between the temporal
variables.  For the couple of hierarchies \eqref{eq23} and
\eqref{eq38}, the relations between vector fields of time variables
are given by
\begin{equation}
\frac{\partial}{\partial t_n}-\big(L_2^{n\over 2}\big)_{\geq 2} = \frac{\partial}{\partial\tau_n}-\big(L_{2m}^{n\over2}\big)_{\geq 1}.
\end{equation}
When $n=3$, we have
\begin{align*}
\frac{\partial}{\partial t_3} =& \frac{\partial}{\partial\tau_3}+\big(L_2^{3\over 2}\big)_{\geq 2} -\big(L_{2m}^{3\over2}\big)_{\geq 1}\\
=& \frac{\partial}{\partial\tau_3} + \frac{1}{8}\Big(-2U_{2y} + U_y^2 + (\mathbb{D}_2U_y)(\mathbb{D}_2U) - (\mathbb{D}_2U)(\mathbb{D}_1U)(\mathbb{D}_1\mathbb{D}_2U) + (\mathbb{D}_1U_y)(\mathbb{D}_1U)\Big)\partial_y\\
&\quad +\frac{1}{16}\Big(-2(\mathbb{D}_2U_{2y}) + (\mathbb{D}_2U_y)U_y - (\mathbb{D}_2U)(\mathbb{D}_1\mathbb{D}_2U)^2 + (\mathbb{D}_2U)U_{2y} - (\mathbb{D}_2U)(\mathbb{D}_1U_y)(\mathbb{D}_1U)\\
&\quad\quad\quad + (\mathbb{D}_1U_y)(\mathbb{D}_1\mathbb{D}_2U) - (\mathbb{D}_1U)(\mathbb{D}_1\mathbb{D}_2U_y) - (\mathbb{D}_1U)(\mathbb{D}_1\mathbb{D}_2U)U_y\Big)\mathbb{D}_2\\
&\quad +\frac{1}{16}\Big(-2(\mathbb{D}_1U_{2y}) + (\mathbb{D}_1U_y)U_y - (\mathbb{D}_1U)(\mathbb{D}_1\mathbb{D}_2U)^2 + (\mathbb{D}_1U)U_{2y} - (\mathbb{D}_2U_y)(\mathbb{D}_2U)(\mathbb{D}_1U) \\
&\quad\quad\quad - (\mathbb{D}_2U_y)(\mathbb{D}_1\mathbb{D}_2U) +
(\mathbb{D}_2U)(\mathbb{D}_1\mathbb{D}_2U_y) +
(\mathbb{D}_2U)(\mathbb{D}_1\mathbb{D}_2U)U_y\Big)\mathbb{D}_1.
\end{align*}

\section{Summary and Problems}
With the help of supercomformal transformations, we have established the relations between two $N=2$ supersymmetric KdV equations and
two supersymmetric HD equations, therefore we generalize our results of $N=1$ supersymmetric reciprocal transformations to the $N=2$ case.
 While this is interesting, there are a number of problems to be solved. We list some of them as follows
\begin{itemize}
\item Our construction above relies on the Lax representations, so it would be important to recover the transformations by means of other property such as
 conservation laws.
\item Studying the implications of our transformations is another interesting problem. Indeed, more properties are known for the KdV cases than for
 the HD cases, for instance, bi-Hamiltonian structures have been constructed for the $N=2$ supersymmetric KdV systems, but we know little about
 Hamiltonian structures for the $N=2$ supersymmetric HD equations apart form the following observation. The systems \eqref{hdt1} and \eqref{hd} can be
  reformulated as
    \begin{align*}
     W_{t_1}=\mathbb{B}\frac{\delta H_1}{\delta W},\qquad W_{t_2}=\mathbb{B}\frac{\delta H_2}{\delta W},
    \end{align*}
    where $\mathbb{B}=W^2{\cal D}_1{\cal D}_2\partial_x W^2$ is a Hamiltonian operator and Hamiltonian functionals are given by
    \begin{align*}
     H_1 = \frac{i}{2}\int \ln W\;\mathrm{d}z\mathrm{d}\theta_1\mathrm{d}\theta_2,\quad\mbox{and}\quad H_2 = \frac{1}{8}\int (\mathcal{D}_1W)(\mathcal{D}_2W)W^{-1}\;\mathrm{d}z\mathrm{d}\theta_1\mathrm{d}\theta_2.
    \end{align*}
\item In the Harry Dym case, two known $N=2$ integrable supersymmetric extensions are proved to be linked with two $N=2$ supersymmetric KdV equations, but there are three rather than two such systems. Thus, this fact seems to indicate that one $N=2$ integrable supersymmetric Harry Dym equation is missing.

\end{itemize}
The progress on solving these and other related problems may be reported elsewhere.

\textbf{Acknowledgment} This work is supported by the National
Natural Science Foundation of China (grant numbers: 
10731080, 10971222) and the Fundamental Research Funds for Central
Universities.

%
%
%
%
%

\end{document}